\documentclass[%
 aip,
rsi,%
 amsmath,amssymb,
 reprint,%
]{revtex4-1}

\usepackage{graphicx} 
\usepackage{float}
\usepackage{color}
\usepackage{dcolumn} 
\usepackage{bm} 
\usepackage{siunitx} 

\usepackage[version=4]{mhchem}

\begin{document}

\preprint{AIP/123-QED}

\title[]{Improving the performance of \ce{Ge_2Sb_2Te_5} materials via nickel doping: Towards RF-compatible phase-change devices}

\author{Pengfei Guo}
\email{guop01@udayton.edu}
\affiliation{Department of Electro-Optics and Photonics, University of Dayton, Dayton, Ohio 45469, USA}
\author{Joshua A. Burrow}
\affiliation{Department of Electro-Optics and Photonics, University of Dayton, Dayton, Ohio 45469, USA}

\author{Gary A. Sevison}
\affiliation{Department of Electro-Optics and Photonics, University of Dayton, Dayton, Ohio 45469, USA}
\affiliation{Air Force Research Laboratory, Sensors Directorate, Wright-Patterson AFB, Ohio 45433, USA}

\author{Aditya Sood}
\affiliation{Department of Electrical Engineering, Stanford University,  Stanford, California 94305, USA}
\affiliation{Department of Mechanical Engineering, Stanford University,  Stanford, California 94305, USA}

\author{Mehdi Asheghi}
\affiliation{Department of Mechanical Engineering, Stanford University,  Stanford, California 94305, USA}

\author{Joshua R. Hendrickson}
\affiliation{Air Force Research Laboratory, Sensors Directorate, Wright-Patterson AFB, Ohio 45433, USA}

\author{Kenneth E. Goodson}
\affiliation{Department of Mechanical Engineering, Stanford University,  Stanford, California 94305, USA}

\author{Imad Agha}
\affiliation{Department of Electro-Optics and Photonics, University of Dayton, Dayton, Ohio 45469, USA}
\affiliation{Department of Physics, University of Dayton, Dayton, Ohio 45469, USA}

\author{Andrew Sarangan}
\email{asaragan1@udayton.edu}
\affiliation{Department of Electro-Optics and Photonics, University of Dayton, Dayton, Ohio 45469, USA}

\date{\today}

\begin{abstract}
High-speed electrical switching of \ce{Ge_2Sb_2Te_5} (GST) remains a challenging task due to the large impedance mismatch between the low-conductivity amorphous state and the high-conductivity crystalline state.  
In this letter, we demonstrate an effective doping scheme using nickel to reduce the resistivity contrast between the amorphous and crystalline states by nearly three orders of magnitude.  
Most importantly, our results show that doping produces the desired electrical performance without adversely affecting the film's optical properties. The nickel doping level is approximately 2\% and the lattice structure remains nearly unchanged when compared with undoped-GST. 
The refractive indices at amorphous and crystalline states were obtained using ellipsometry which echoes the results from XRD. The material's thermal transport properties are measured using time-domain thermoreflectance (TDTR), showing no change upon doping. 
The advantages of this doping system will open up new opportunities for designing electrically reconfigurable high speed optical elements in the near-infrared spectrum.

\end{abstract}

\maketitle

\setlength{\parindent}{4ex}


Amorphous-to-crystalline reversible phase transitions are found in many materials, most notably in chalcogenide glasses.  
This phase change, which is typically induced by heat, either using an electrical current or optical excitation, results in very large changes in optical and electrical properties. These properties have been exploited in nonvolatile data storage devices such as rewritable digital versatile disks (DVD-RWs) and phase change random access memories (PCRAMs)\cite{yi2003novel,cho2005highly,lee2007gesbte}. 
More recently, phase-change materials have been proposed as means for phase and amplitude modulation of light \cite{metasurface,Hendrickson}, exploiting the large index contrast between the amorphous and crystalline states.  
Among the various chalcogenide alloys, the pseudo-binary \ce{Ge_2Sb_2Te_5} (GST) has been widely studied because of its fast crystallization speed, reversible phase transition and good endurance over a wide temperature range. 
However, there are two primary challenges that limit applications of GST in electronics, especially at high speeds such as in radio-frequency (RF)-compatible devices.  
The first is the extremely large resistivity in the amorphous state (several hundred $\Omega$-m). 
This necessitates either a large voltage, or a long (tens of ns) pulse,  to dissipate sufficient power in the film to induce the phase transition . 
The second challenge is the large impedance difference between the amorphous and crystalline phases. 
In order to maintain a constant load impedance (nominally 50 $\Omega$), one would have to utilize an impedance matching circuit with fast tunability, rendering the system nearly impractical for RF applications.
Most of the work on modifying the GST electrical properties has so far
been focused on reducing the switching current by increasing the resistivity of the GST in the amorphous state. 
Unfortunately, this also significantly increases the required voltage/switching time, which is especially important for quasi-static applications such as PCRAMs\cite{PCRAMspeed}.

During the past few years, different dopant elements have been explored for modifying the electrical and optical properties of GST as well as for improving its long-term stability. 
Previous work includes nitrogen \cite{seo2000investigation,NdopedGSTShelby}, carbon\cite{CdopedGSTZhou,li2018carbon}, oxygen\cite{Privitera2004b}, aluminum\cite{Guoqiang2012ImprovedMemory}, silver\cite{Zhou2014}, gold\cite{AudopedGSTSingh}, titanium\cite{Wei2011},  tungsten\cite{Guo2014a}, and copper \cite{Ding2014StudyMemory}.  
Although these studies showed that the resistivity of GST can be modified by adding dopants into the material, none have demonstrated the ability to reduce the resistivity of the amorphous state of the host GST material.  Zhu \textit{et. al.} \cite{zhu2015ni} studied nickel-doped GST (hereafter, GST-Ni) for high speed phase change memory applications and improved data retention. 
Although that work shows that GST-Ni has a lower resistivity in the amorphous state and a higher resistivity in the crystalline state compared to pure GST, the optical properties and thermal conductivities were not examined, which are critical for GST-based high-speed optical applications.

In this letter, we focus on nickel-doped GST films fabricated via  co-sputtering.  
We examine a number of different properties of pure GST and 2\% GST-Ni using energy dispersive X-ray spectroscopy (EDX), Raman spectroscopy, X-ray diffraction (XRD), time-domain thermoreflectance (TDTR), and ellipsometry.  
The aim of this work is to develop a nickel-doped GST platform suitable for high speed switching.  Specifically, the goal is to demonstrate a path towards minimizing the impedance mismatch between the amorphous and crystalline states without compromising the usefulness of the material in optical devices.  
To the best of our knowledge, this is the first work that comprehensively studies the Ni-doping of GST and its effects on electrical, thermal and optical properties.


Nickel-doped GST samples were prepared by co-sputtering using dual 3-inch circular magnetron cathodes.  During deposition, one cathode was supplied from a 13.56 MHz RF source (for GST) and another cathode supplied by a DC source (for nickel).  
Prior to deposition, the substrates (silica glass for electrical, XRD, Raman and ellipsometry characterization, and crystalline silicon for TDTR) were cleaned using
Acetone - Methanol - Isopropyl alcohol (AMI) to remove organic contaminations, and dried with nitrogen gas. The substrates were then placed in the  chamber and pumped down to a base pressure below 1 $\mu$Torr. 
%
%

\begin{table}[h]
\begin{ruledtabular}
\begin{tabular}{ccc}
DC Power (W) & Pressure (mT) & Concentration (\%)\\
\hline
300 & 8 & $\approx 20$\\
150 & 8 & $\approx 15$\\
75 & 8 & $\approx 12$\\
75 & 6 & $\approx 10$\\
75 (with mesh) & 6 & $\approx 2$\\
\end{tabular}
\end{ruledtabular}
  \caption{Deposition conditions and nickel concentration in GST-Ni films.}
  \label{table:sputter}
\end{table}

The RF discharge power on the GST target was maintained at 100 W while the DC power on the nickel target was set at various levels, ranging from 300 W down to 75 W to reduce the nickel concentration in the deposited films.  
As shown in Table \ref{table:sputter}, reducing the DC power on the nickel target had the expected effect of reducing the doping level. Concentrations were obtained by averaging the measured EDX compositions from three separate areas on each sample.  
However, since the power cannot be arbitrarily scaled down to very low levels without compromising the stability and repeatability of the plasma, the smallest doping concentration achievable was around 10\%.  
Therefore, a wire mesh mask was implemented to further reduce the nickel deposition rate.  
This mesh, placed between the substrate and the sputter target, helped reduce the deposition rate by condensing a fraction of the incident nickel atoms on the mesh. 
Using the co-sputtering system in combination with this meshed shutter, we were able to reduce the concentration down to 2\% (or even lower by using a finer mesh). 



\begin{figure}[ht]
\includegraphics[width=\linewidth]{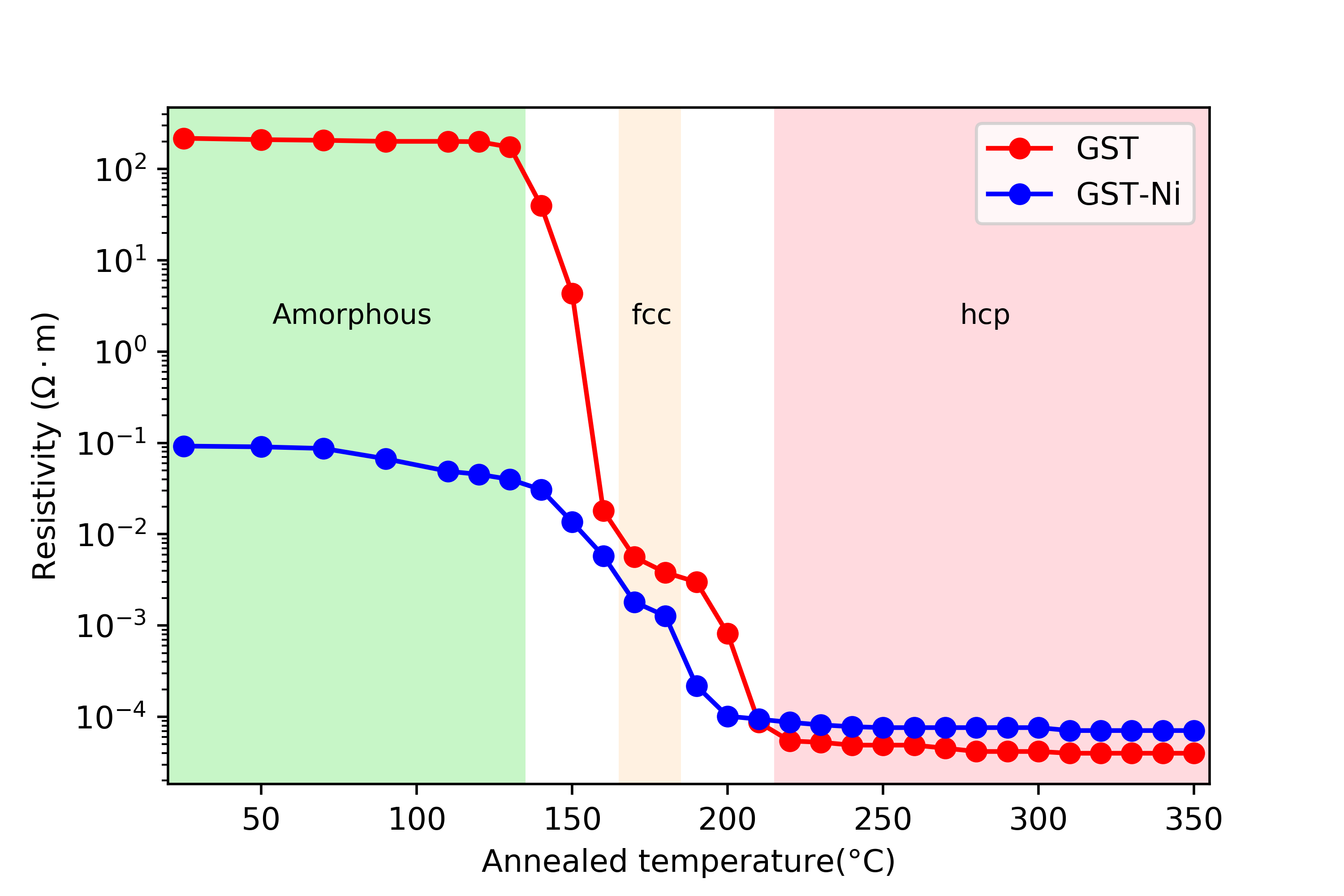}\\
\caption{Relationship between resistivity and annealing temperature of the as-deposited pure GST and GST-Ni films.}
\label{fig:resistivity}
\end{figure}

The deposited film thickness was obtained by measuring the step height using a stylus profiler. 
The thickness of the films used for the electrical, Raman, XRD, and ellipsometry experiments was approximately 200 nm, while that for TDTR was 405 nm.
The resistivity of the films as function of annealed temperatures were obtained by first annealing each GST and GST-Ni samples at a series of discrete temperatures from \SI{25}{\celsius} to \SI{350}{\celsius} and then measuring their sheet resistance values by using a four-point probe. 
%
%
The annealing process was performed by placing the film in direct contact with the hot plate (i.e. film facing down towards the hot plate) in ambient conditions for 3 minutes and returning to room temperature for the measurement.

The resistivity curves of GST (red) and GST-Ni (blue) as a function of annealing temperature are shown in Fig. \ref{fig:resistivity}. The resistivity of the GST films shows two discrete transitions.  
The first transition near \SI{150}{\celsius} corresponds to the amorphous to face-centered-cubic (fcc) phase transition, while the second transition near \SI{230}{\celsius} is due to the fcc to hexagonal close-packed (hcp) phase\cite{kato2005electronic}.  
On the other hand, the amorphous state resistivity of the GST-Ni film is lower than the undoped GST film by more than 3 orders of magnitude, while the resistivity in the crystalline states have comparable resistivity values (which is already highly conductive and hence, doping has little effect). 

In order to explain the decrease in resistivity of the amorphous state, we note that the atomic radius of Ni (135 pm) is close to that of Te (140 pm)\cite{Guo2014a}, which makes it likely that  nickel atoms  act as substitutional impurities in the crystal lattice of GST.  
%
%
In fact, similar substitutional effects of tungsten\cite{Guo2014a} and indium\cite{lazarenko2016influence} in doped GST have been previously reported. As such, it is likely that nickel acts an acceptor for germanium, making it a p-type dopant. 
To verify this hypothesis, we measured the electrical polarity of the films using the hot probe method \cite{hotprobe}.  
The undoped GST films exhibited a p-type behavior, which is consistent with the report from Kato \textit{et. al.}\cite{kato2005electronic}.  
Even after nickel-doping the films remained dominantly p-type.  
Furthermore, since the doped GST films exhibited a higher conductivity than the undoped GST films, this suggests that nickel atoms act as additional acceptor sites in the GST crystal lattice. 
%
%


\begin{figure}[ht]
\centering
\includegraphics[width=\linewidth]{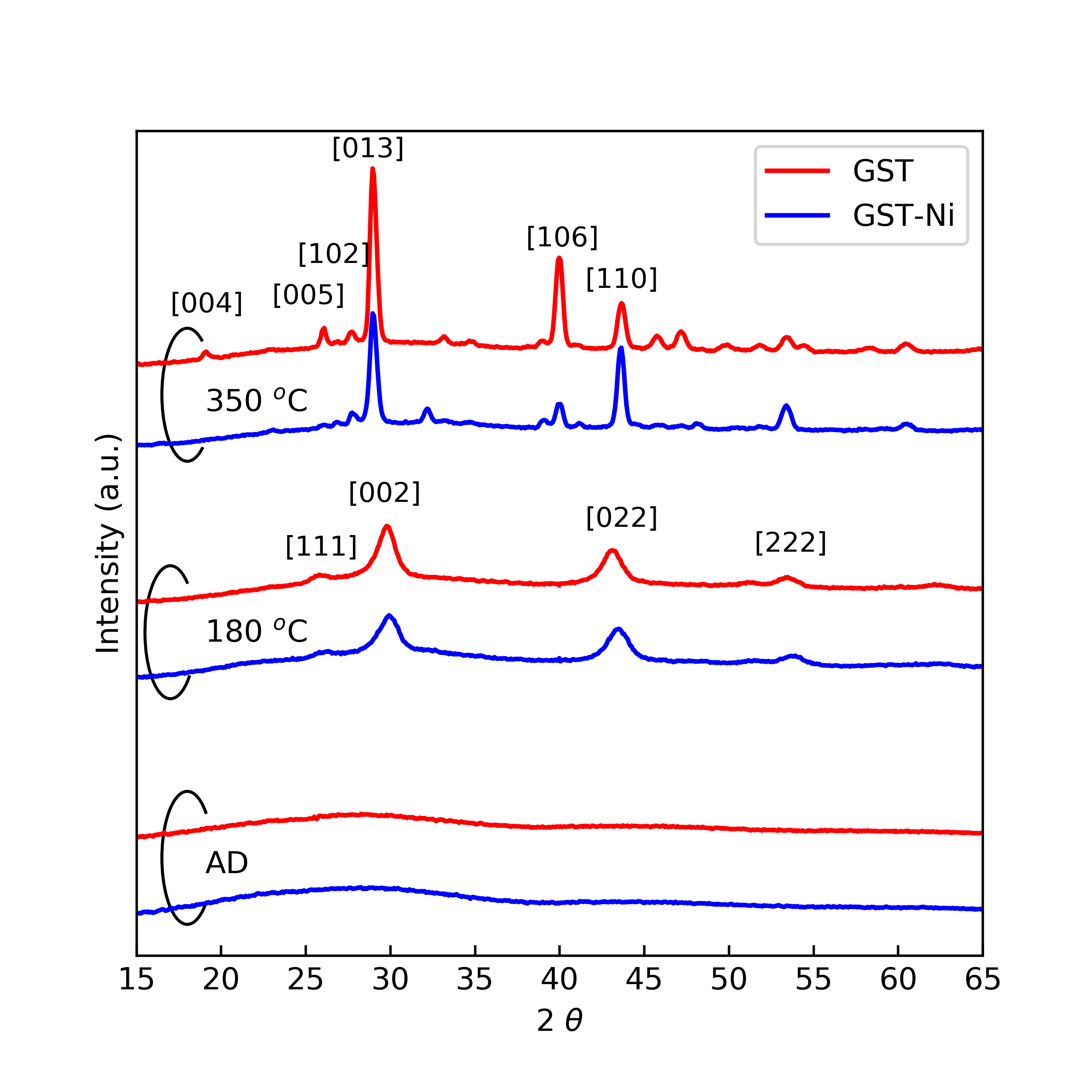}\\
\caption{XRD spectra of GST and GST-Ni samples for As-Deposited state (bottom) and after annealing at \SI{180}{\celsius} (middle), \SI{350}{\celsius} (top). The curves have been offset for clarity.} 
\label{fig:xrd_3in1}
\end{figure}

The structural information of the films related to their crystallization behaviors were characterized by XRD.
The scanning range of the diffraction angles is from \SI{15}{\degree} to \SI{65}{\degree}.  Fig.~\ref{fig:xrd_3in1} shows the XRD patterns of the as-deposited GST and GST-Ni films after annealing them at \SI{180}{\celsius}, \SI{250}{\celsius} and \SI{350}{\celsius}, respectively. 
Since there are no identifiable peaks in the XRD patterns of the as-deposited films, we confirm that all as-deposited films are in the amorphous state. 
The undulance around \SI{29}{\degree} may be due to the [200] peak \cite{Wei2007}.
After annealing the samples at \SI{180}{\celsius}, both the GST and GST-Ni films crystallized into the metastable fcc state shown by the XRD peaks [111], [002], [022], and [222], which are typical for the NaCl like cubic cell structure\cite{Sarangan2018}.
After both samples are annealed at \SI{350}{\celsius}, new XRD peaks appear that correspond to the hexagonal state\cite{matsunaga2004structures}.
The Ni-doped GST shows lower XRD peak intensities, which indicates a lower crystallinity compared to pure GST. 
However, the diffraction peaks of both the samples are well aligned,  which confirms that the small amount of nickel (2\% in this case) does not significantly change the lattice structure of the host material. 

\begin{figure}[ht]
\includegraphics[width=\linewidth]{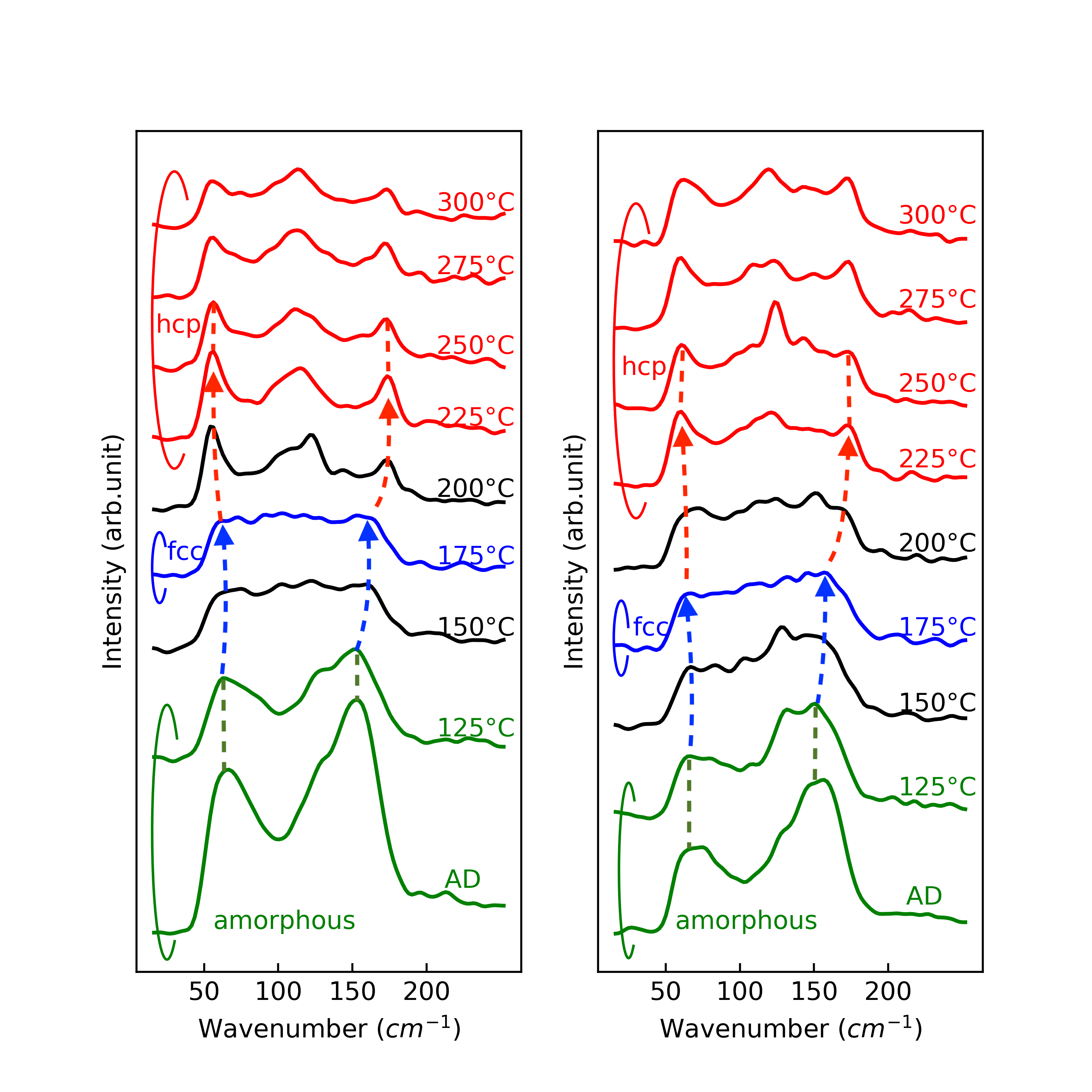}\\
\caption{Temperature dependence of Raman spectra for GST (left), GST-Ni (right) from as-deposited amorphous state to crystalline state after annealing at \SI{300}{\celsius}.}
\label{fig:raman}
\end{figure}

To investigate the impact of nickel-doping on the short-range order of the films, we acquired the Raman spectra from these samples after annealing them in a series of temperatures between \SI{25}{\celsius} and \SI{300}{\celsius}.
Fig. \ref{fig:raman} shows the measured data between the spectral range of 25  cm$^{-1}$ and 250  cm$^{-1}$ for the undoped-GST films (left) and the doped GST-Ni films (right). 
At around \SI{150}{\celsius} annealing temperature, coincident with the amorphous-to-fcc transition, the peak near 66 cm$^{-1}$ experiences a blue shift, while the peak near 150 cm$^{-1}$ experiences a red shift.
Beyond \SI{225}{\celsius}, the spectra does not show any discernible changes at higher annealing temperatures, indicating that the film is fully crystallized and stabilized in the hexagonal state.  
This is also echoed by the results from the resistivity variance plot in Fig. \ref{fig:resistivity}.  The black curves in Fig. \ref{fig:raman} indicate mixed transition states. The lower black line is a transition state between the amorphous and fcc states, and the upper black line is between the fcc and hcp states. 
The assignment of different Raman peaks can be inferred from previous reports\cite{jang2010origin,sosso2011raman,nvemec2012amorphous}.    
Most importantly, our results show that the Raman spectra of GST and GST-Ni have similar characteristics throughout the entire phase transition temperature range, suggesting that there are no dramatic changes to the GST crystal matrix as a result of the nickel doping. 


\begin{figure}[ht]
\centering
  \includegraphics[width=\linewidth]{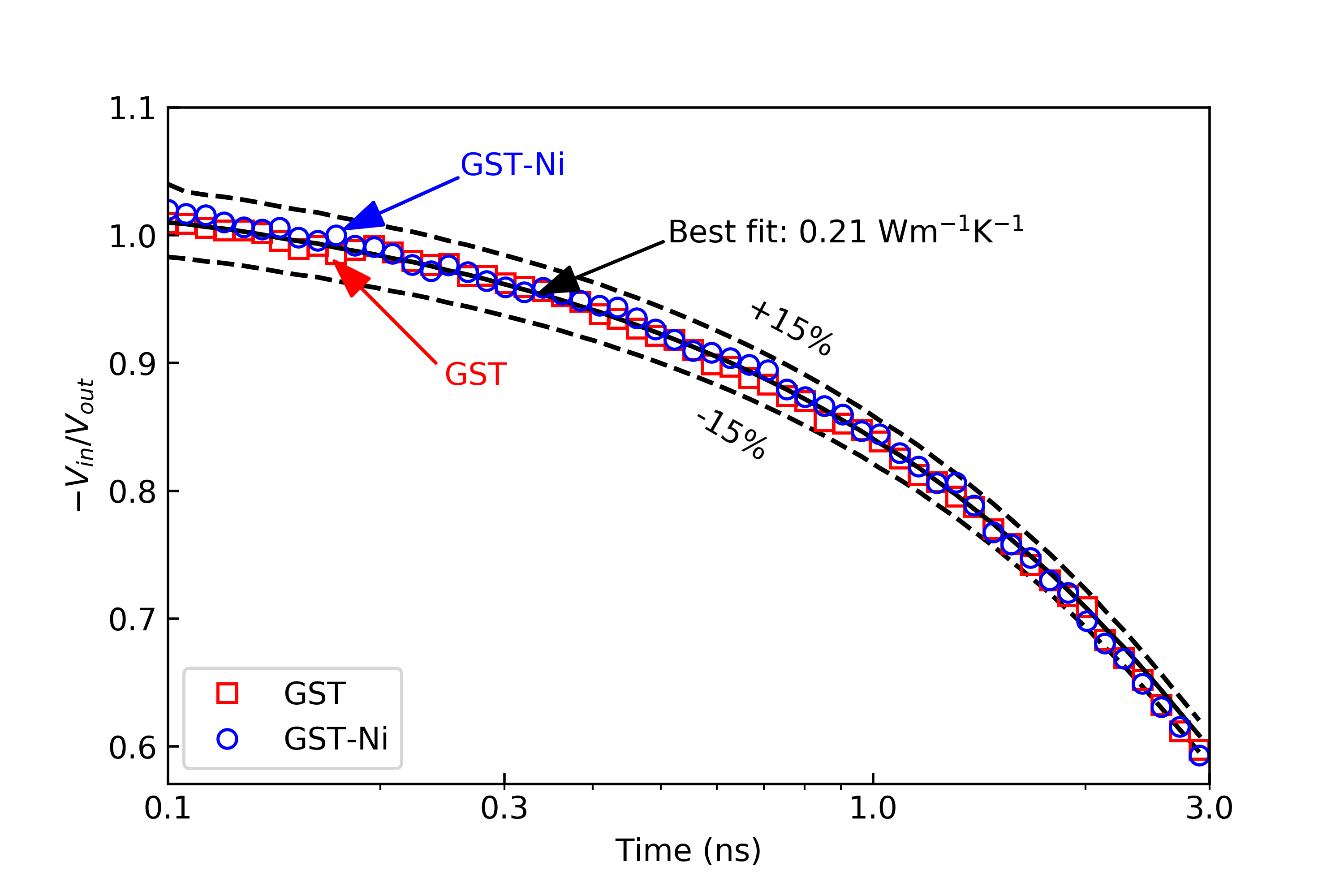}\\
  \caption{
Representative TDTR data for GST (red squares) and GST-Ni (blue circles) films. The black solid line denotes the best fit thermal conductivity of 0.21 Wm$^{-1}$K$^{-1}$. Black dashed lines correspond to $\pm 15\%$ error bounds.}
  \label{fig:tdtr}
\end{figure}

Since phase-change is a primarily thermal phenomenon, it is important to understand the effect of doping on the thermal transport characteristics of the 
switching material. To investigate the impact of Ni doping on the thermal conductivity of GST, we performed time-domain thermoreflectance (TDTR) measurements. 
TDTR is an ultrafast optical pump-probe technique that is used to measure the thermophysical properties of thin films down to a few tens of nanometers in thickness.  
Details of this method, and of our setup are provided elsewhere\citep{ahn2015energy,sood2016anisotropic}.  
Briefly, 9 ps pump pulses (532 nm wavelength) are used to heat the surface of the sample (405 nm thick GST on crystalline Si) through optical absorption within a thin 85 nm Al transducer layer. 
Time-delayed probe pulses (1064 nm wavelength), arriving 100 ps to 3 ns after the pump pulses, measure the rate of cooling of the metal layer as heat diffuses downwards into the GST film and Si substrate, by monitoring changes in its reflectivity. 
These data are fit to the solution of a 3D heat diffusion model to extract the unknown thermophysical properties of interest within the film stack. 
The properties of the Al layer and Si substrate are measured independently or taken from literature. 
The measurement is largely insensitive to the thermal boundary conductance at the Al/GST and GST/Si interfaces; each is fixed at 100 MWm$^{-2}$K$^{-1}$ during the fitting. 
The volumetric heat capacity of GST is taken to be 1.4 Jcm$^{-3}$K$^{-1}$ based on literature estimates\citep{zalden2014specific}, and assumed to be the same for the GST-Ni sample. 
%
%

In these measurements, we used a root mean square spot size (1/e$^{2}$ diameter) of 8.8 $\mu$m, and optical powers of 11 and 3 mW for pump and probe, respectively. The pump pulses were amplitude modulated at 10 MHz for lock-in detection. Representative TDTR time-decay curves for the GST and GST-Ni samples are shown in Fig. \ref{fig:tdtr}. From these, we extract a thermal conductivity of 0.21$\pm 0.03$ Wm$^{-1}$K$^{-1}$ for both samples, which is 
in good agreement with literature for amorphous GST \citep{ahn2015energy, lyeo2006thermal}.
Importantly, we find that within the experimental uncertainties of the measurement, there is no detectable change in the thermal conductivity of GST due to the 2\% Ni doping.

For optical device applications of GST, such as switchable on-chip  spatial light modulators, the most important parameter is the refractive index contrast between the amorphous and crystalline states of GST.
If doping is utilized to reduce the resistivity contrast to enable better electrical impedance matching, it is important to verify that this is not done at the expense of the refractive index contrast between the two states.
To examine this, spectroscopic ellipsometry was used to study the refractive index and extinction coefficients of GST and GST-Ni thin films in both the amorphous and hcp crystalline states.  
%
%

\begin{figure}[ht]
   \includegraphics[width=\linewidth]{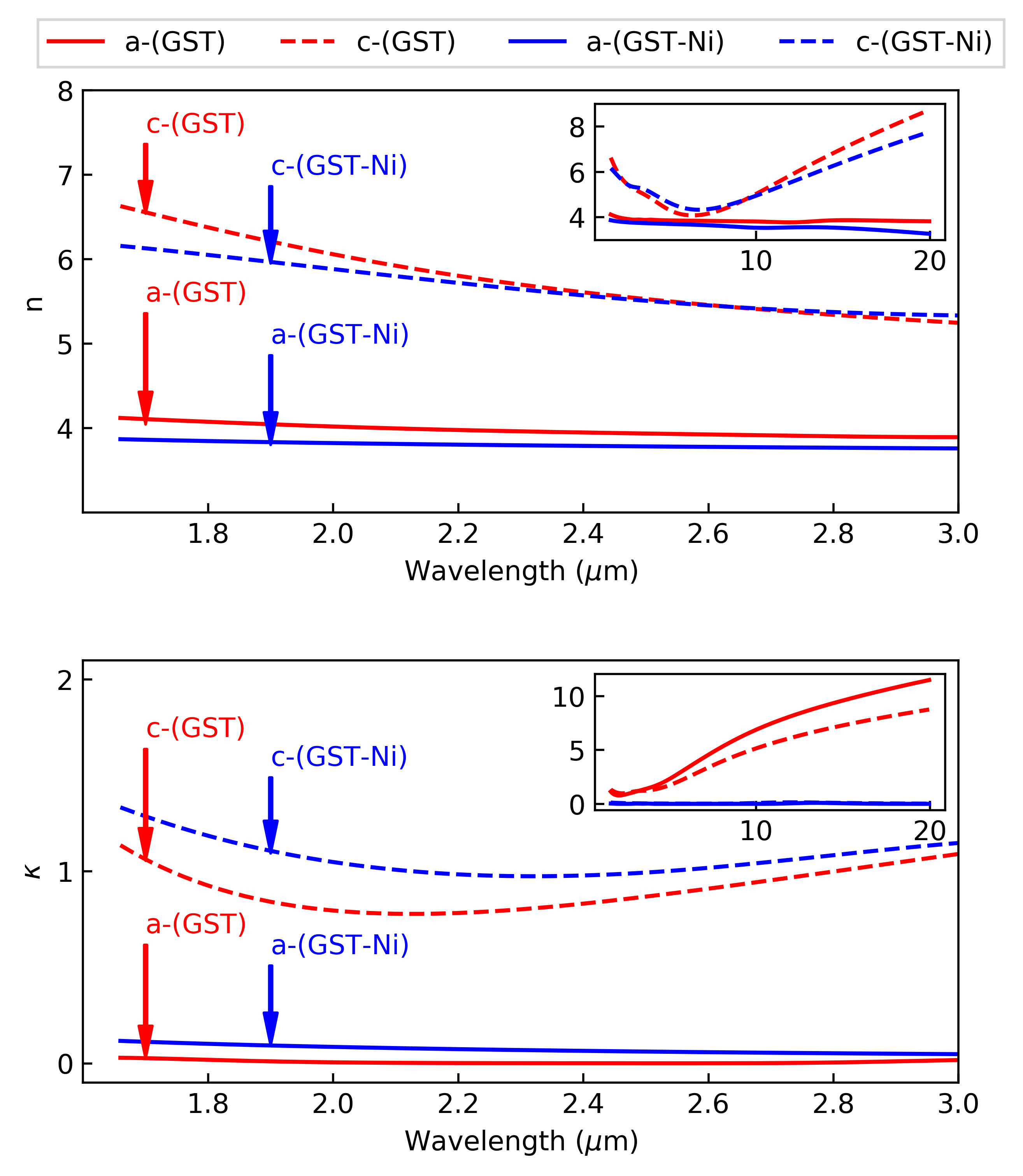}
  \caption{Measured refractive index (top) and extinction coefficient (bottom) of GST and GST-Ni film in amorphous  and crystalline states. The insets show the $n$ and $\kappa$ in the spectrum up to 20 $\mu$m. 
}
  \label{fig:nk_ac}
\end{figure}

The results are plotted in Fig. \ref{fig:nk_ac}. The undoped GST is shown by red lines, and the nickel-doped GST are shown by the blue lines.
%
%
The solid curves represent the amorphous state (indicated by a-) and the dashed curves represent the hcp crystalline state (indicated by c-).
In the wavelength range of interest, the refractive index $n$ of both GST and GST-Ni increases by about 2 after the amorphous-to-crystalline transition, while the extinction coefficient $\kappa$ increases by about 1, which is consistent with prior results\cite{Sarangan2018}.
The inset plots show the same data over a much larger spectral range.
We can conclude from this data that nickel doping has not adversely affected the optical constants of GST, despite the dramatically altered electrical resistivity.


In summary, in this letter, we have carried out a comprehensive study of the effects of nickel doping in GST films pertaining to their electrical, structural, thermal, and optical properties. 
The resistivity of GST-Ni in the amorphous state was reduced by over 3 orders of magnitude as compared to the undoped GST, while the resistivity of the crystalline states remained nearly the same.
This enables the fabrication of devices that can be readily impedance matched to high speed electronics irrespective of the actual GST state. 
The XRD and Raman spectroscopy results shows the nickel dopant does not significantly change the lattice structures of the GST host material in both the amorphous and crystalline states. Furthermore, TDTR measurements indicate that thermal properties are nearly identical pre and post-doping. 
More importantly, our ellipsometry results demonstrate that the refractive index contrast is well preserved, and nickel doping does not adversely affect the optical constants in either the amorphous or crystalline states. 
These results indicate that GST-Ni is a promising candidate for high-speed RF-compatible electrically switchable optical devices.
\\
\\ 
\indent This work is supported by the National Science Foundation under Grant Nos. $1710273$ and $1709200$, and the University of Dayton graduate school. The authors thank Rachel H. Rai from the Department of Chemical and Materials Engineering at the University of Dayton for the Raman measurements, and Heungdong Kwon from the Department of Mechanical Engineering at Stanford University for inputs on TDTR measurements.


\bibliography{aipsamp.bib}

\end{document}